\documentclass[useAMS,usenatbib]{mn2e}

\usepackage{graphicx}

\title[Fast X-ray variability of RU\,Peg]{Resolving different sources of fast X-ray variability of the dwarf nova RU\,Peg in quiescence}
\author[A. Dobrotka, S. Mineshige and J.-U. Ness]{
A. Dobrotka$^{1,2,3}$\thanks{E-mail: andrej.dobrotka@stuba.sk},
S. Mineshige$^2$\thanks{E-mail: shm@kusastro.kyoto-u.ac.jp}
and J.-U. Ness$^3$\thanks{E-mail: juness@sciops.esa.int}\\
$^1$Slovak University of Technology in Bratislava, Faculty of Materials Science and Technology in Trnava, Institute of Materials\\
Science, Paul\'inska 16, 91724 Trnava, Slovak Republic\\
$^2$Department of Astronomy, Graduate School of Science, Kyoto University, Sakyo-ku, Kyoto 606-8502, Japan\\
$^3$European Space Astronomy Center, PO Box 78, 28691 Villanueva de la Ca\~nada, Madrid, Spain\\
}

\begin{document}

\date{Accepted ???. Received ???; in original form \today}

\pagerange{\pageref{firstpage}--\pageref{lastpage}} \pubyear{2012}

\maketitle

\label{firstpage}

\begin{abstract}
We analysed an X-ray light curve of the dwarf nova RU\,Peg taken by XMM-Newton with a duration of 46300\,s. The power density spectrum has a complicated shape with two red noise and two white noise components, indicating the presence of two turbulent regions. We developed a statistical "toy model" to study light curves with variability produced by an unstable turbulent accretion flow from the inner disc. Our results are consistent with a disc truncation radius maximally $0.8 \times 10^9$\,cm. We found that any fluctuation of the viscous mass accretion at the inner disc are visible as UV and X-ray variations with the same break frequency in the power density spectrum. This process is generating low frequency variability. A second break suggests the presence of a faster X-ray variability component which must be generated by another process likely localised between the inner disc and the white dwarf.
\end{abstract}


\section{Introduction}
\label{introduction}

Dwarf novae are a subclass of a group of semidetached interacting binaries called cataclysmic variables (see \citealt{warner1995} for review). In these systems the secondary star filling its Roche lobe is losing matter which flows towards the white dwarf primary. In the case of weak magnetic fields of the primary, an accretion disc is formed. The inner disc is interacting with the slowly rotating white dwarf primary in a small region called boundary layer. Furthermore, dwarf novae are known for variability caused by the viscous-thermal instability caused by changes in the ionisation state of hydrogen while the matter flows through the disc (\citealt{hoshi1979}, \citealt{meyer1981}, \citealt{lasota2001}). This variability is characterised by two basic states, i.e. high state (outburst) with higher mass accretion rates through the disc and low state (quiescence) with a low mass accretion rate (see \citealt{osaki1974} for the original idea).

X-rays in cataclysmic variables are thought to originate from the boundary layer (see \citealt{frank1992}, \citealt{kuulkers2006} for review). The matter at the inner disc is rotating with supersonic Keplerian velocity. When it enters the boundary layer, it decelerates to reach the white dwarf surface angular velocity, yielding loss of kinetic energy that is then emitted as X-rays. Approximately half of the gravitational potential energy is radiated from the boundary layer. The other half is radiated from the disc through viscous processes. At low accretion rates, the boundary layer is optically thin and cools inefficiently. Therefore, it reaches high temperatures and emits hard X-rays. This occurs in dwarf novae in quiescence. During an outburst, the mass accretion rate rises and the gas becomes optically thick. It cools more efficiently and emits soft X-rays and EUV. A steady flow of matter through the boundary layer results in a steady energy release, and observed rapidly variable X-ray light curves thus indicate an unstable accretion flow.

The light curves of cataclysmic variables show stochastic variability also in other wavelengths as presented in, e.g., \cite{bruch1992}, \cite{yonehara1997}, \cite{bruch2000}, \cite{baptista2004}, \cite{baptista2008}, \cite{dobrotka2010}, \cite{dobrotka2012}. Variable energy release at all energy scales is expected because the accretion flow from the secondary is turbulent through the entire disc down to the white dwarf. Variability is produced within the disc itself or at its outer edge where it interacts with the matter stream from the secondary. The X-ray fluctuations coming from the boundary layer are poorly studied, mainly because of the high cost of deep X-ray observations that are required to yield enough photon statistics for high time resolution variability studies. For example, the brighter low mass and high mass X-ray binaries that contain neutron stars or black hole primaries, providing higher X-ray flux, are much better studied in X-rays (see e.g. \citealt{vanderklis2005}).

RU\,Peg is a system with an orbital period 8.99 hours, a white dwarf with mass $1.29_{-0.20}^{+0.16}$\,M$_{\odot}$ and secondary with mass $0.94 \pm 0.04$\,M$_{\odot}$ (\citealt{stover1981}, \citealt{wade1982}, \citealt{shafter1983}). The distance $282 \pm 20$\,pc was derived by \citet{johnson2003} with Hubble Fine Guidance Sensor. The system has been studied in X-ray by \citet{balman2011}. Based on measured luminosity the authors derived a mass accretion rate $2 \times 10^{-11}$\,M$_{\odot}$\,yr$^{-1}$ assuming the primary mass 1.3\,M$_{\odot}$. Cross-correlation analysis of the light curve yielded the peak at 0\,s delay to be asymmetric, where the asymmetry was well fitted with an additional delayed component when allowing for a time lag of 97 - 109\,s (\citealt{balman2012}). The X-ray variability patterns generating the asymmetry lagged behind the UV.

In this article, we search for physical processes responsible for unstable accretion flow in the boundary layer in the dwarf nova RU\,Peg. We reanalyse the data set used by \citet{balman2012} with a more sophistacted approach. While the authors found a single break frequency in the power density spectrum and identified it with the orbital frequency at the inner edge of the disc, our more detailed analysis suggests an alternative interpretation of the break frequencies. In Sect.~\ref{observations} we present the studied X-ray light curve taken with the EPIC detectors on board XMM-Newton. Observations and data reduction are described in Sect.~\ref{pds_analysis}. Subsequently synthetic light curves are modelled and their characteristics are compared to the observed parameters in Sect.~\ref{simulation_study}. The results are discussed and summarised in Sect.~\ref{discussion} and \ref{summary}, respectively.

\section{Observations}
\label{observations}

\begin{figure}
\includegraphics[width=59mm,angle=-90]{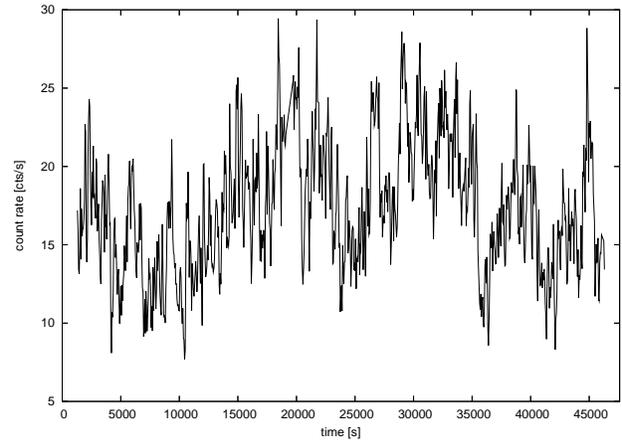}
\caption{X-ray light curve of RU\,Peg. The three light curves from XMM-Newton detectors MOS1, MOS2 and PN have been summed up.}
\label{light_curve}
\end{figure}

While high signal-to-noise spectra can always be accumulated by long exposure times, high resolution in time requires the most sensitive instruments. We therefore use data from the most sensitive X-ray mission XMM-Newton, already analysed by \citet{balman2011} and \citet{balman2012}. We used archival data of RU\,Peg, ObsID 0551920101, taken June 9, 2008. The data were downloaded from the XMM-Newton Science Archive, and science products obtained with the Science Analysis Software (SAS), version 11.0. We used the tool {\tt xmmextractor} to re-generate calibrated events files from which in turn light curves were extracted for all instruments. Optimised extraction regions for the EPIC detectors were calculated with the tool {\tt eregionanalyse} while for the Reflection Grating Spectrometer (RGS) and the optical monitor (OM), standard extraction regions were used by {\tt xmmextractor}. We used the RGS and OM light curves only for consistency checks while for EPIC, we combined the pn and MOS light curves for best signal-to-noise for detailed analysis. The combined EPIC light curve is shown in Fig.~\ref{light_curve}.

\section{Power density spectra}
\label{pds_analysis}

We first reproduce the results by \citet{balman2012} using the full light curve in a single analysis step. We then pursue a statistical approach by break the entire light curve into several shorter pieces and perform power density spectra (PDS) analysis on each one of them. Subsequently we determined a mean PDS and measured the PDS parameters. This approach yields smaller scatter and thus more accurate conclusions. We note that we use different units for the y-axis as described below.

\subsection{Determination of Parameters}
\label{pds_parameters}

\begin{figure}
\includegraphics[width=125mm,angle=-90]{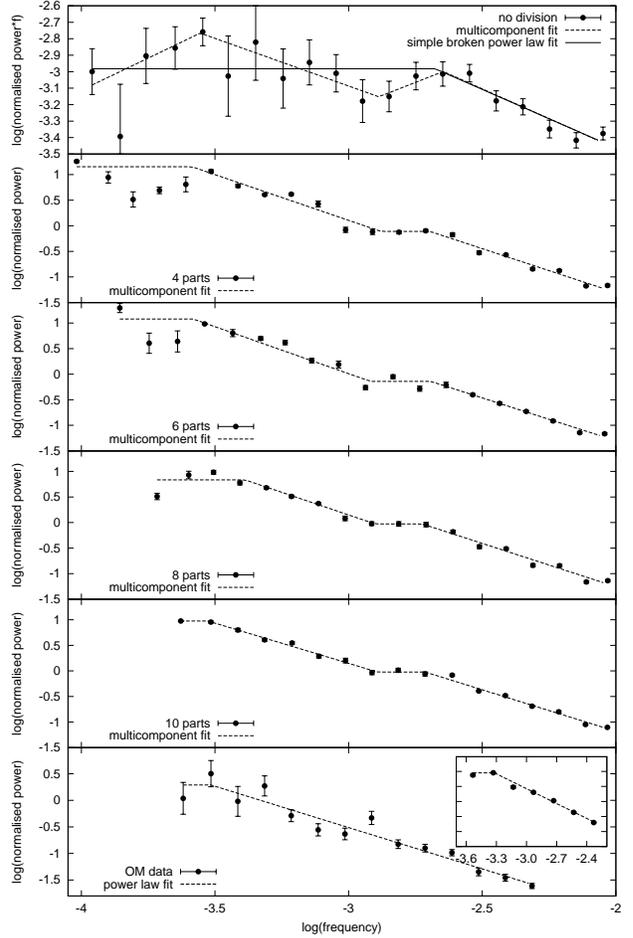}
\caption{Binned PDS of the X-ray light curve divided into 4, 6, 8, and 10 segments, and PDS from the OM data (see legends in each panel). The main PDS is from the undivided light curve and the inset panel is from the light curve divided into 10 parts. The errors are those of the mean, while the solid lines represent the power law fits.}
\label{pds_observed}
\end{figure}

For timing analysis of the summed XMM-Newton EPIC light curves, we applied the Lomb-Scargle algorithm (\citealt{scargle1982}). We first used the full data set to calculate a power density spectrum (PDS). In the top panel of Fig.~\ref{pds_observed}, we show this PDS in the same way as \citet{balman2012}, i.e. with the y-axis in units of power $p$ multiplied by frequency $f$. We binned the PDS into 0.1dex in the logarithmic frequency scale and find consistent results as \citet{balman2012}. The main difference is that we show the low frequency part with higher resolution. This part of the PDS appears flat and noisy. Considerably higher binning is required to decrease the scatter (as used by \citet{balman2012}) but such procedure obscures many details that we intend to resolve, thus our different approach.

The scatter is an inherent characteristics of the PDS which can not be reduced by using higher time resolution or longer duration of the light curve. One way to reduce the scatter is to calculate a sample of PDS from different segments of the entire light curve from which a mean PDS is determined. This is a standard procedure in the case of optical ground-based observations from different nights. We sub-divided the total 46.3-ks light curve into 4, 6, 8, and 10 subsamples and calculated the mean PDS; a larger number of subsamples reduces the range of useful frequencies in the PDS at the low frequency end. The high frequency end is constrained by the time resolution of the light curve. A higher resolution extends the PDS to higher frequencies where the PDS is dominated by white noise. The transition to white noise is not abrupt, and we found 50\,s binning to be the optimal resolution.

Furthermore, following \citet{papadakis1993}, the resulting mean PDS is binned and each bin is evaluated as $\langle$log $p \rangle$ rather than log$\langle p \rangle$. Therefore, as units for the y-axis, we used $p$ instead of $f \times p$\footnote{It is worth to note, that a flat PDS in units of $f \times p$ is a power law in PDS in units of $p$.} that was used by \citet{balman2012}. For any additional fitting, the binned $\langle$log $p \rangle$ values are used.

Fig.~\ref{pds_observed} (except the top panel) shows the new PDS binned into 0.1dex in the logarithmic frequency scale. The reduction in scatter can clearly be seen and can be attributed to the use of more light curves per mean PDS, but mainly to different y-axis units. The PDS shape is robust against the choice of sub dividing the light curve. The low frequency PDS is characterised by a "plateau" while the high frequency part decreases in power with increasing frequencies. This shape is typical for observed flickering, i.e. white noise for the low frequency end and red noise at high frequencies. This PDS is usually fitted by a constant and a power law, but in our PDS, an additional constant "plateau" is present between frequency values -3.0 and -2.5dex. Such PDS shape has frequently been observed in X-ray binaries (see e.g. \citealt{sunyaev2000}, \citealt{vanderklis2005}, \citealt{done2007}). Therefore, we fitted the binned PDS with a multicomponent model, i.e. white noise - red noise - white noise - red noise (constant - power law - constant - power law). This more complex model yields considerably better reduced $\chi^2$ of 6.29 over a simple white noise - red noise model with reduced $\chi^2$ equal to 11.70 in the 10 subsample case. Fig.~\ref{pds_observed} shows the multicomponent fits (in the top panel we show also the equivalent version with four power laws for the different y-axis units) for every light curve division and in Table~\ref{pds_fit_parameters} we list the corresponding fitted parameters. The results are very similar except for the first power law index and first break frequency. The uncertainty studies (see below) show that these parameters have the largest error and all values agree within the estimated uncertainty. A considerably higher number of the light curve subsamples would be needed to improve the results requiring a much longer observation to keep reasonable subsample duration. With the existing data, we thus continue our analysis with 10 subsamples and the corresponding fitted parameters.
\begin{table}
\caption{Fitted PDS parameters calculated from the summed light curve from the three XMM-Newton detectors MOS1, MOS2, PN (marked as X) and from OM (marked as UV). $n$ is the number of light curve subsamples, $pl_1$, $pl_2$ are the power low indices (low and high frequency ends, respectively) and $f_1$, $f_2$ and $f_3$ are the break frequencies. The values in parentheses are 1-$\sigma$ uncertainty for $n = 10$ derived in Sect~\ref{pds_uncertainty}.}
\begin{center}
\begin{tabular}{lcccccr}
\hline
\hline
$n$ & $pl_1$ & $pl_2$ & log($f_1$) & log($f_2$) & log($f_3$) &\\
\hline
4 & -1.79 & -1.72 & -3.59 & -2.88 & -2.70 & X\\
6 & -1.84 & -1.67 & -3.58 & -2.92 & -2.69 & X\\
8 & -1.77 & -1.70 & -3.39 & -2.90 & -2.72 & X\\
10 & -1.57 & -1.62 & -3.53 & -2.89 & -2.71 & X\\
10 & ($\pm$0.77) & ($\pm$0.37) & ($\pm$0.25) & ($\pm$0.17) & ($\pm$0.13) & X\\
1 & -1.56 & -- & -3.52 & -- & -- & UV\\
10 & -1.65 & -- & -3.32 & -- & -- & UV\\
10 & ($\pm$0.54) & -- & ($\pm$0.26) & -- & -- & UV\\
\hline
\end{tabular}
\end{center}
\label{pds_fit_parameters}
\end{table}

While the main objective of this paper is the study of X-ray variability, we also compute a PDS for the OM light curve that we will need for comparison with the X-ray PDS in Sect.~\ref{discussion}. The data are of lower quality, with lower resolution, and the light curve is not continuous. Therefore, they are not suitable for a detailed analysis as for the X-ray data. The PDS derived from the undivided light curve with frequency range and resolution as the previous 10 light curve X-ray case is shown in the bottom panel of Fig.~\ref{pds_observed}. The inset panel shows the PDS after the same procedure was applied as for the X-ray light curve using the 10 light curve subsample per mean PDS. The PDS parameters are summarised in Table~\ref{pds_fit_parameters}.

\subsection{Uncertainty estimate}
\label{pds_uncertainty}

\begin{figure}
\includegraphics[width=90mm,angle=-90]{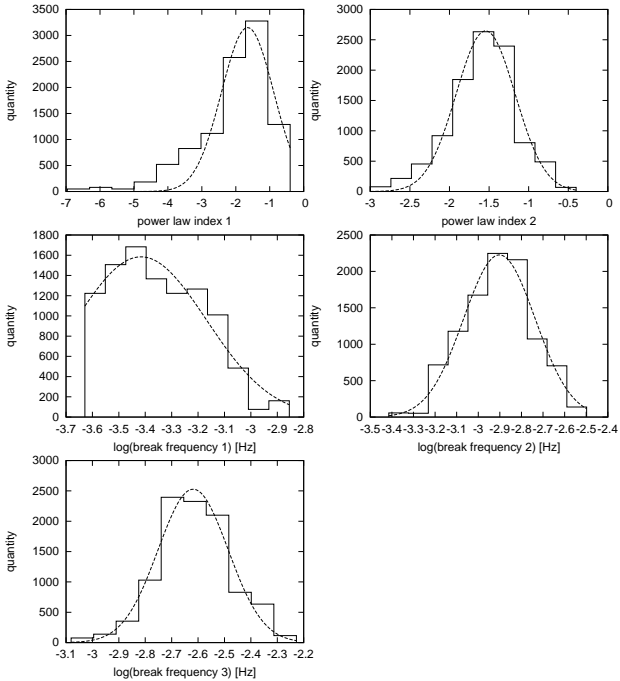}
\caption{Histograms of the fitted PDS parameters for an estimate of uncertainties. Ten light curves were produced to generate a mean PDS. The simulations were repeated 10000 times. Dotted lines are the Gaussian fits to the histograms.}
\label{scatter_multi}
\end{figure}

To assess an uncertainty in our statistical approach, we took into account the number of light curves used per mean PDS as it determines the scatter. We simulated 10 light curves with the same duration and sampling as the observed data and we calculated the mean PDS as in Sect.~\ref{pds_parameters}. Variability was modelled following the measured PDS shape using the method of \citet{timmer1995}. As the frequencies in the inverse Fourier transform are discrete, the break frequencies were chosen to be close to that observed. We repeated the process 10000 times and a Gaussian function was fitted to the histograms of fitted parameters (Fig.~\ref{scatter_multi}). We define the uncertainty of parameters by the 1-$\sigma$ values of the Gaussian fits (Table~\ref{pds_fit_parameters}).

The histograms do not always show a perfect Gaussian profile, i.e. in the first break frequency. This is owing to the duration of the light curve subsample. The low frequency end prevents the histogram from completing the Gaussian shape. This is more problematic in the first power law index case. The distribution is asymmetric, hence the fitted 1-$\sigma$ uncertainty can be underestimated. But even with this underestimated value we can already conclude that the uncertainty is rather large (49\,\%) and this will be taken into account in our later PDS modelling.

For the UV data, we performed the same uncertainty estimate. We are again interested in the mean PDS with $\langle$log $p \rangle$ as y-axis value, hence we searched for the uncertainty of the 10 light curve division case. The values are given in Table~\ref{pds_fit_parameters}. Taking this 1-$\sigma$ value from a Gaussian fit as error, the break frequency in the UV PDS agrees with the $f_1$ value from the X-ray PDS and not with $f_3$ as suggested by \citet{balman2012}. However, we determined the 1-$\sigma$ (67\%) errors while \citet{balman2012} used 2-$\sigma$ (95\%) errors. If we expand our error range to the same level, we get the same results as \citet{balman2012}. However, this approach is counterproductive, because any different values would agree within the errors when increasing the error range, which also increases the confidence error range\footnote{Typical examples of comparable PDS from different energies are obvious in the case of intermediate polars V1223\,Sgr (\citealt{revnivtsev2010}) and EX\,Hya where the PDS slopes and break frequencies are almost equal (\citealt{revnivtsev2011}).}. Therefore, when testing whether two values agree, we prefer to operate at a higher level of general doubt in exchange for smaller error ranges. Finally, we suggest that the UV frequency $f_1$ is different from the X-ray frequency $f_3$ with a minimal confidence level of 67\%, but the UV $f_1$ agrees well with the X-ray frequency $f_1$.

\section{Simulation study}
\label{simulation_study}

In order to test possible physical origins of the observed fast variability patterns, we simulated synthetic light curves based on given physical assumptions, performed the same PDS analysis as described above, and compared the resulting parameters to the ones derived from the observations.

\subsection{The model}
\label{kh_instability}

\begin{figure}
\includegraphics[width=85mm,angle=0]{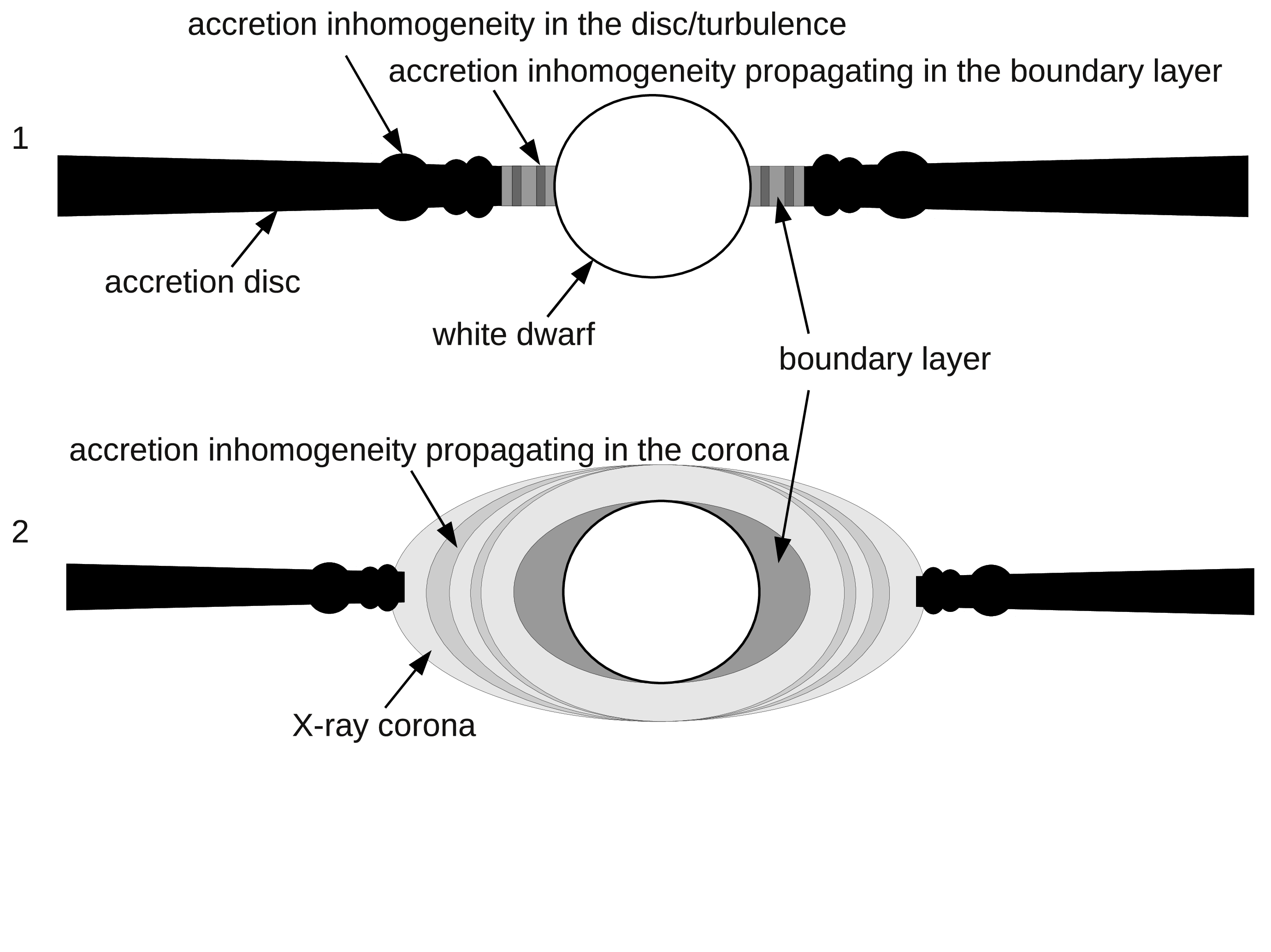}
\caption{Schematic (not to scale) pictures of investigated/discussed models. See text for details.}
\label{ilustration_1}
\end{figure}

It is believed that the disc is highly turbulent (\citealt{dobrotka2010}, \citealt{dobrotka2012}, \citealt{romanova2012}) and the mass flow through the disc is not steady. Therefore, also the flow from the inner disc to the boundary layer should have a similar turbulent behaviour. This process was simulated by \citet{romanova2012}. The authors found that MRI-driven (magnetorotational instability, \citealt{balbus1998}) turbulence develops in the disc and mass accretion rate variability at the central star surface is associated with accretion of individual turbulent cells. This turbulent inhomogeneous flow is producing light curve modulation. In other words, a fluctuation in mass inflow from the inner disc can increase or decrease the overall X-ray flux (situation 1 in Fig~\ref{ilustration_1}). This can be observed as variability on viscous time scales. In the following we study variability generated by turbulent unstable mass supply from the inner disc. This mechanism was also proposed by \citet{revnivtsev2010}.

\subsection{Light curve modelling}
\label{modelling_KH}

Our method of light curve simulation is based on a simple idea of a fragmented accretion flow similar to \citet{dobrotka2010} and \citet{dobrotka2012}. The matter is "falling" from the inner disc toward the white dwarf in a flow that consists of blobs of matter with dimension scale $x$. This scale is exponentially distributed following the function
\begin{equation}
f(x) = {\rm exp}(ax).
\label{exp_funkcia}
\end{equation}
where $a < 0$. The boundary conditions of the distribution are $f(0)=1$ and $f(H)=0.01$, where $H$ is the scale height of the inner disc as maximum dimension scale $x_{\rm max}$. The equation (\ref{exp_funkcia}) becomes
\begin{equation}
f(x) = {\rm exp} \left[ \frac{{\rm ln}~f(H)}{H}~x \right].
\label{exp_funkcia_2}
\end{equation}
The scale height of the inner disc is estimated from
\begin{equation}
H = c_{\rm s} \left( \frac{r}{GM} \right)^{1/2} r,
\label{Hdisc}
\end{equation}
(see e.g. \citealt{frank1992}) where $r$ is the distance from the centre, $G$ is the gravitational constant, $M$ is the primary mass and $c_{\rm s}$ is the sound speed calculated from
\begin{equation}
c_{\rm s} = \left( \frac{k T}{\mu m_{\rm H}} \right)^{1/2},
\label{cs}
\end{equation}
where $k$ is the Boltzmann constant, $\mu$ is the mean molecular weight (0.615), $m_{\rm H}$ is the hydrogen atom mass and $T$ is the temperature at the inner disc. Following the disc instability model (\citealt{lasota2001}) this temperature in quiescent novae discs should be below the value of hydrogen ionisation, i.e. less than 8000\,K. Observations of V4140\,Sgr yield a temperature of 6000\,K in the inner disc (\citealt{borges2005}), whereas observations of V2051\,Oph also in quiescence showed a temperature of the inner disc to be 13500\,K (\citealt{vrielmann2002}). We base our modelling on the theory of the disc instability model and we use an inner disc temperature of 5000\,K. Additional tests of our simulations showed that the results do not change significantly when assuming temperatures of order $\pm 10^3$\,K and we consider them robust.

In our algorithm, we chose a certain number of bodies $N_{\rm b}$ with a dimension scale $x$ (a sphere with a radius $x$ in our approximation) following the distribution function (\ref{exp_funkcia_2}). The total mass is calculated using the density at the inner disc. Dividing this total mass by the measured mass accretion rate of RU\,Peg, we get the total duration of the synthetic light curve. Therefore, the chosen initial number of bodies $N_{\rm b}$ should give the final light curve duration similar to the observed light curve, i.e. 4630\,s in the 10 light curve subsample division (Sect.~\ref{pds_parameters}) or if the duration is longer, only the 4630\,s part is taken for a subsequent analysis. Every spherical blob of matter flowing from the inner disc releases an amount of energy and produces a sine square shape flare\footnote{The instantaneous energy release of a spherical body penetrating into the boundary layer or stellar surface is proportional to the surface of a spherical section with diameter $\rho = x\,sin(\beta)$, and the surface is proportional to $\rho^2$. The angle $\beta$ is the phase of the penetration; $\beta = 0$ and $\beta = \pi$ are the first and last contacts (beginning and end of the flare) respectively. We also tried sine and triangular shaped flares and it has little impact on results.}. All flares with certain duration and amplitude (calculation explained later) are randomly redistributed into the calculated duration. A synthetic light curve sampled as the observed one is constructed.

The duration of every sine square flare is calculated from the body dimension scale $x$ divided by the radial viscous velocity $v_{\rm r}$
\begin{equation}
v_{\rm r} = \frac{3 \nu}{2 r} \left( 1 - \sqrt{\frac{r_{\rm wd}}{r}} \right)^{-1},
\label{visc_time_scale}
\end{equation}
where $r_{\rm wd}$ is the white dwarf radius estimated from \citet{nauenberg1972} ($r_{\rm wd} = 2.98 \times 10^8$\,cm) and $r$ is the distance from the centre. The parameter $\nu$ is the turbulent kinematic viscosity parametrised by a dimensionless coefficient $\alpha$ as (\citealt{shakura1973})
\begin{equation}
\nu = \alpha c_{\rm s} H,
\label{viscous_coeff}
\end{equation}
where $H$ is the scale height of the inner disc and $c_{\rm s}$ is the sound speed. Typical values of $\alpha$ for dwarf novae in quiescence are of order 0.01 (see \citealt{lasota2001} for review). We used an interval from 0.01 to 0.05.

We caution that our model is not a physical model, but rather a statistical approach. We are not calculating the real flare amplitude in flux, but only a relative amplitude as a function of the flare duration which is crucial in timing analysis. The energy $E$ released by the spherical body with a dimension scale $x$, density $\rho$ in the distance $r$ can be expressed as
\begin{equation}
E \sim G \frac{M~4/3~\pi x^3 \rho}{r} \propto x^3.
\end{equation}
The flare has a sine square shape and the released energy is proportional to the flare surface $S$. This sine square surface is equal to the triangle surface with the same amplitude and duration. Therefore, the surface $S$ can be expressed by the simple analytic formula
\begin{equation}
S = \frac{t A}{2},
\end{equation}
where $t$ is the flare duration and $A$ is the amplitude that we seek to determine. Using $E \sim S$ and $t \simeq x/v_{\rm r}$ we get
\begin{equation}
A \propto x^2 v_{\rm r}~~{\rm or}~~A = K x^2 v_{\rm r}.
\end{equation}
$K$ is a constant which has to be calculated. We choose a relative amplitude of 1 when the body dimension scale is equal to the scale height of the inner disc, which is satisfied by the condition $1 = K H^2 v_{\rm r}$. Therefore, the final amplitude is calculated as
\begin{equation}
A = \left( \frac{x}{H} \right)^2.
\end{equation}

\begin{figure}
\includegraphics[width=84mm,angle=-90]{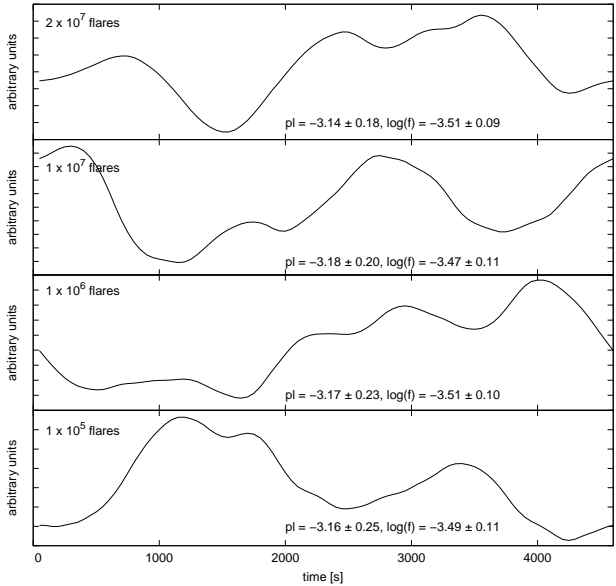}
\caption{Simulated light curves with $2.0 \times r_{\rm wd}$ as inner disc radius and $\alpha=0.05$ in equation~\ref{viscous_coeff} testing different numbers of flares (see text for details). The labels at the bottom of each panel are the mean power law index ($pl$) and logarithm of break frequency (log($f$)) with 1-$\sigma$ uncertainty obtained after 1000 simulations.}
\label{lc_simulated_visc}
\end{figure}

In Fig.~\ref{lc_simulated_visc} we show four examples of simulated light curves assuming $2.0 \times r_{\rm wd}$ as inner disc radius $r_{\rm in}$ and $\alpha=0.05$ in equation~\ref{viscous_coeff}. These artificial light curves were used to test the influence of the number of flares per investigated light curve duration. As already mentioned in this section, the number of flares is selected following the mass transfer rate, but sometimes the calculations are very time consuming. Therefore, we investigated whether a reduction in the number of flares has an effect on the resulting PDS. The labels in Fig.~\ref{lc_simulated_visc} show the mean PDS parameters with 1-$\sigma$ uncertainty calculated from 1000 simulated light curves. It is clear that the total number of flares is not important. Crucial is the relative quantity of different time scales. Therefore, later when necessary we used a lower number of flares than $N_{\rm b}$, while keeping the same light curve duration.

\subsection{Results}
\label{results_turbul}

\begin{figure}
\includegraphics[width=82mm,angle=-90]{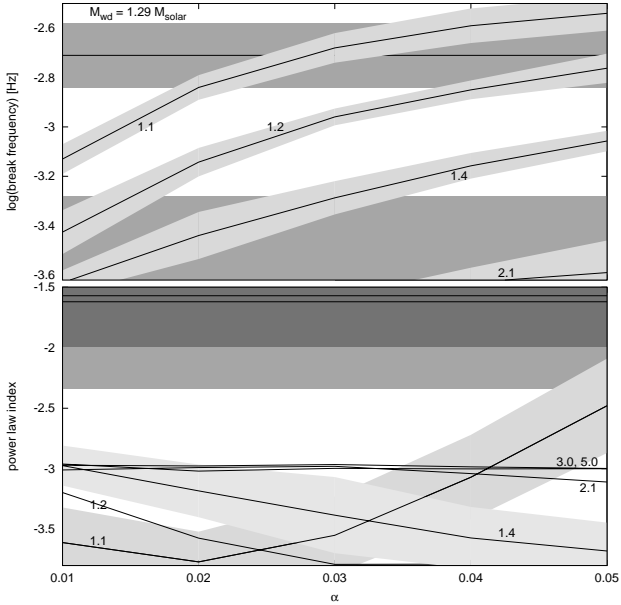}
\caption{Probability diagram of simulated break frequency (top) and power law index (bottom) as a function of the disc parameter $\alpha$ for different models assuming a white dwarf mass of 1.29\,M$_{\rm \odot}$ and four inner disc radii, in units of white dwarf radius, given near the corresponding curves. 1-$\sigma$ confidence intervals are marked with light shadings around the corresponding curves, where in the bottom panel, some confidence ranges are omitted for readability. The observed values of break frequency and power law index with their 1-$\sigma$ uncertainties (see Table~\ref{pds_fit_parameters}) are included with horizontal dark shaded areas. Intersections between dark and light shadings in top and bottom panels indicate values of $\alpha$ that are consistent with the observations.}
\label{results_pl_ctf_visc}
\end{figure}

The typical shape of the resulting PDS is a broken power law. We calculated 10 light curves per mean PDS (as performed with observed data in Sect.~\ref{pds_parameters}) and fitted a broken power law model to the resulting mean PDS. We repeated the process 1000 times. We determine power law index and break frequency with uncertainties from a Gaussian function fitted to the respective PDS parameter histograms. We investigated different models with different inner disc radii. The mean value with a 1-$\sigma$ uncertainty as a function of $\alpha$ is depicted in Fig.~\ref{results_pl_ctf_visc}. The observed PDS parameters with 1-$\sigma$ intervals from Table~\ref{pds_fit_parameters} are shown as darker shaded areas. We searched for solutions matching (one of) two observed broken power laws with parameters $f_1$, $pl_1$ or $f_3$, $pl_2$.

The modelled break frequencies agree with the observed $f_1$ for all tested models with approximately $r_{\rm in} = 1.2 - 2.1 \times r_{\rm wd}$. Models with larger radii do not show this break frequency in the studied interval. For these larger inner disc radii models, the break frequency exceeds the lower end of the calculated PDS defined by the observation duration. The PDS has no broken power law shape in the studied frequency interval, but it is a simple power law. Those models are therefore not adequate. Furthermore, the higher observed break frequency $f_3$ is satisfied only for very small inner disc truncations, $1.2 \times r_{\rm wd}$ or less.

Finally, all modelled power law indices except the model with $r_{\rm in} = 1.1 \times r_{\rm wd}$ are considerably below the observed parameter intervals. At large inner disc radius ($r_{\rm in} = 2.1 - 6.0 \times r_{\rm wd}$), the resulting power law slopes are insensitive against variations in studied $\alpha$ interval. In fact, when increasing $\alpha$, the power law index curve passes through a minimum and then starts to increase. The minimum reaches lower values of $\alpha$ for smaller inner disc radii which can be seen in the model with $r_{\rm in} = 1.1 \times r_{\rm wd}$. This model marginally matches the observed power law $pl_1$, but does not match the corresponding frequency $f_1$.

Apparently, our modelling is able to reproduce the observed break frequency, but not the slope (or the corresponding slope). What is the real meaning of this red noise slope? The PDS power is strongly decreasing (with a power law index about -3). This means that variability patterns that are considerably shorter than the break frequency are not present or significant (Fig.~\ref{lc_simulated_visc}). This means that there must be an additional process adding shorter variability to the observed light curve with frequencies higher than the break frequency. Our simulations have shown that the viscous scenario is not able to reproduce this short term variability and a different physical mechanism is required to explain this part of the PDS.

\subsubsection{Impact of uncertainty in white dwarf mass}

Our modelling depends on the primary mass which has considerable uncertainty with a range of values 1.09 to 1.45\,M$_{\odot}$ (\citealt{stover1981}, \citealt{wade1982}, \citealt{shafter1983}). We tested also these extreme values\footnote{The upper value is above the Chandrasekhar mass limit and is thus unconstrained. Using the uppermost limit 1.45\,M$_{\odot}$ would result in a negative number using \citet{nauenberg1972} radius estimate and 0 radius using value 1.44\,M$_{\odot}$ (useful for Equation~\ref{visc_time_scale}). Therefore, as an upper value we used 1.40\,M$_{\odot}$ with "secure distance" from the critical mass.}. In these models, we keep the inner disc radius constant in cm units, because $r_{\rm wd}$ changes with mass. All subsequent inner disc radii are given in $r_{\rm wd}$ units for better comparison with Section~\ref{results_turbul}.

The simulated PDS slopes show the same behaviour as in the previous simulations. The power law indices of models satisfying break frequency $f_1$ or close to it remain considerably below the observed values. Once the simulated break frequency approaches the observed value $f_3$, the power law index approaches the observed higher values as well. The two parameters are therefore correlated and, as concluded in the previous section, we are able to model the observed break frequency, but not the slope (or the corresponding slope), implying that an additional process is generating the fast variability.

\begin{figure}
\includegraphics[width=82mm,angle=-90]{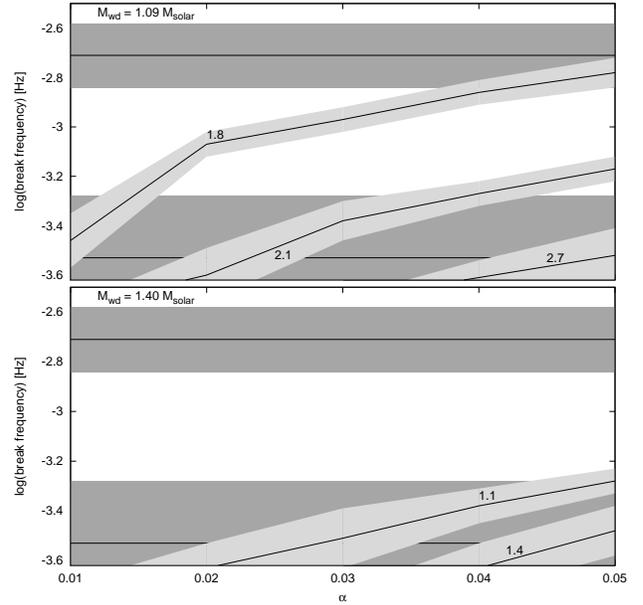}
\caption{The same as top panel of Fig.~\ref{results_pl_ctf_visc} assuming white dwarf masses 1.09\,M$_{\rm \odot}$ (top) and 1.40\,M$_{\rm \odot}$ (bottom), exploring the uncertainty range.}
\label{results_pl_ctf_visc_mass_var}
\end{figure}

In general, a higher primary mass implies a lower break frequency and vice versa. In Fig.~\ref{results_pl_ctf_visc_mass_var} we illustrate models for the lower\footnote{The lower mass limit yields a larger white dwarf radius of $0.48 \times 10^9$\,cm. Therefore, the models with $r_{\rm in} \leq 1.4 \times r_{\rm wd}$ from Fig.~\ref{results_pl_ctf_visc} make no sense in this case.} and higher mass limit. The lower mass moves the simulated curves toward higher break frequencies, while the higher mass did the opposite. The maximal possible inner disc radius for the lower and upper mass case is approximately 2.7 and $1.4 \times r_{\rm wd}$, respectively.

As already mentioned, all models have power laws below the observed values except the model with $r_{\rm in} = 1.2 \times r_{\rm wd}$ for $M = 1.09$\,M$_{\rm \odot}$. This solution matches the high frequency part of the observed PDS, i.e. broken power law with break frequency $f_3$ and power law index $pl_2$.

\section{Discussion}
\label{discussion}

A typical shape of a flickering PDS in cataclysmic variables is a broken power law, i.e. white noise and red noise in the low- and high frequency parts, respectively. The observed PDS of the dwarf nova RU\,Peg in quiescence has two such components. Therefore, the goal is to explain two break frequencies $f_1$ and $f_3$ (Tab.~\ref{pds_fit_parameters}). Our simulations of fast X-ray variability are based on unstable mass accretion from the inner disc to the boundary layer. However, with our model we are able to explain only one break frequency. An additional mechanism is needed to explain the rest of the PDS. In general the brightness variability is caused by a fragmented accretion flow that originates in different regions of the accretion flow. The dimension scale of the fragments and radial velocity influence the PDS characteristics. In general, smaller and/or faster fragments generate shorter time scale events. This moves the break frequency towards higher values.

\subsection{Identification of the break frequencies}

Due to large uncertainty of the measured break frequency $f_1$ and the large interval of tested values of $\alpha$, all investigated models with inner disc radii approximately about $2 \times r_{\rm wd}$ and below match the lower break frequency $f_1$. When simulating the accretion process with very small inner disc radius (lower than $1.2 \times r_{\rm wd}$), the simulated values match also the higher break frequency $f_3$. More detailed discussion is needed to assign the correct frequency to the modeled viscous process.

\subsubsection{Lower break frequency $f_1$}
\label{lower_break_frequency}

A disc with inner disc radius lower than $1.2 \times r_{\rm wd}$ has a boundary layer radial width $r_{\rm bl}$ lover than $0.2 \times r_{\rm wd}$. Our observations were taken during a phase of quiescence, hence the boundary layer was in the optically thin regime. The transition from the optically thick disc to the optically thin boundary layer is abrupt. The temperature can exceed $10^8$\,K and the gas density drops to $10^{-9}$\,g\,cm$^{-3}$ close to the white dwarf surface for a mass accretion rate of $2 \times 10^{-11}$\,M$_{\odot}$\,yr$^{-1}$ (\citealt{narayan1993}, \citealt{medvedev2002})\footnote{The distance uncertainty of 20\,pc yields a mass accretion rate uncertainty of $0.3 \times 10^{-11}$\,M$_{\odot}$\,yr$^{-1}$.}. The radial width and temperature of the boundary layer increase with decreasing mass accretion rate. For the mass accretion rate of RU\,Peg this width is approximately 0.4 times the white dwarf radius (\citealt{narayan1993}). However, taking the break frequency $f_3$ as reference, the boundary layer radial width upper limit is only $0.2 \times r_{\rm wd}$ (Fig.~\ref{results_pl_ctf_visc}). Therefore, not in agreement with the estimate of \citet{narayan1993}. The model with $r_{\rm in} = r_{\rm bl} = 1.4 \times r_{\rm wd}$ agrees only with the break frequency $f_1$. The same is valid also using the lower primary mass limit. We can not study the upper limit case, because the upper mass limit is not constrained and the simulated PDS parameters are very sensible to mass changes close to the Chandrasekhar mass limit. Therefore, our unstable mass accretion rate model supports rather the break frequency $f_1$ as the observational manifestation of this viscous turbulent process. 

Furthermore, any fluctuations in mass accretion rate through the inner disc must modulate in the same way both the UV radiation of the inner disc and X-ray of the boundary layer. Therefore, both types of variability must have the same characteristic frequencies and must be correlated. In the Section~\ref{pds_analysis} we identified also a break frequency in UV PDS. This break frequency is in agreement with the X-ray break frequency $f_1$ and do not agree within the errors with the second detected break frequency $f_3$. Furthermore, \citet{balman2012} studied the correlation between X-ray and UV data and found that the data are correlated. The strongest correlation is at $\sim 0$\,s delay. The cross-correlation function is asymmetric with an additional component in the delayed part. This indicates that the X-rays are partly delayed with respect to the UV variations. The delay is smooth with no visible discrete dominant peak. But an additional fit centred around 97-109\,s explains this asymmetry well. Following the interpretation of \citet{balman2012} this means, that part of the X-rays are delayed by $\sim 100$\,s. Following the authors the measured time delay is the time travel from the inner disc to the stellar surface. Alternative explanation can be in different flare profiles in the two bands as in Cyg X-1 (\citealt{negoro2001}). The time lag between soft and hard X-rays is misleading in this black hole binary and it represents slower growth of the soft X-ray radiation. The cross-correlation function do not show any significant peak suggesting the real time lag (\citealt{maccarone2000}). Instead, the peak is at zero lag with obvious asymmetry. The larger the energy difference of two compared bands, the stronger the asymmetry. In RU\,Peg case an inhomogeneity/blob of matter is formed in the inner disc and starts to radiate in UV (the slow rise of UV radiation). When it enters the boundary layer it starts to radiate in X-ray via free-free radiation with much shorter time scale (rapid rise of X-ray). At this transition boundary, where the matter blob enters the boundary layer, the UV is generated practically at the same time as X-rays (culmination of both wavebands with no time lag). Subsequent evolution of the flare is driven by the decreasing viscous flow\footnote{The only driving velocity is the viscous velocity of the flow from the disc. The radial velocity in the corona or the boundary layer does not play a role here.} of matter from the blob at the inner disc (slow UV and X-ray decline). Therefore, the observed asymmetry of the cross-correlation function can be due to accretion flow generating different radiative processes with different time scales at the disc-boundary layer or disc-corona (see Section~\ref{corona} for discussion of the corona) transition.

Therefore, mainly because of the break frequency agreement in both studied wavebands, we conclude that the frequency generated by the viscous process is the frequency $f_1$. The idea of the mass flow through the inner disc modulating both UV and X-ray with the same break frequency was proposed also by \citet{revnivtsev2010}, but with different values (the difference is explained in Section~\ref{pds_uncertainty}).

\subsubsection{Higher break frequency $f_3$}
\label{higher_brek_frequency}

The second break frequency $f_3$ is generated by a faster process than the viscous mass accretion rate. This suggests that the generating region is localised closer toward the central star than the inner disc radius. The boundary layer is a possible candidate. The angular velocity decreases rapidly within the boundary layer until the surface angular velocity of the white dwarf is reached. This region must be subject to Kelvin-Helmholtz instability. This instability occurs whenever there is a velocity gradient to the flow along the direction that separates two fluids (see \citealt{padmanabhan2000} for review). The accretion flow supplying the boundary layer is fragmented into small blobs which produce unstable X-ray emission, i.e. studied fast variability (situation 3 in Fig~\ref{ilustration_2}). Such hydrodynamical processes are poorly studied and the precise determination of the break frequency $f_3$ via this mechanism is beyond the scope of this paper. However, few time scale estimates can give us a rough image of the possible mechanisms.

\citet{balman2012} already applied a Keplerian time scale assuming that the frequency $f_3$ is produced at the inner disc. The resulting inner disc radius from our measurement of $f_3$ is $1.04 \times 10^{10}$\,cm, or $35 \times r_{\rm wd}$. This radius is much higher than the upper limit derived in this work (Fig.~\ref{results_pl_ctf_visc}). Therefore, this possibility is not in agreement with our simulations.

When the blob of matter enters the boundary layer, it is heated to the temperature of about $10^8$\,K and due to low densities it is radiating via free-free emission. The frequency integrated emissivity $J$ is given by
\begin{equation}
4 \pi J = 1.4 \times 10^{-27} g\,n^2 T^{1/2}\,[{\rm erg\,cm^{-3}\,s^{-1}}],
\end{equation}
where $n$ is the number density approximated as density $10^{-9}$\,g\,cm$^{-3}$ divided by the hydrogen/proton mass ($n \simeq 6 \times 10^{14}$\,cm$^{-3}$), $T$ is the plasma temperature $10^8$\,K and $g$ is the frequency average of the Gaunt factor varying between 1.1 and 1.5 (see e.g. \citealt{padmanabhan2000}). The cooling time scale is estimated as thermal energy per unit volume of the gas ($\sim 3 n k T$, $k$ is the Boltzmann constant) divided by the emissivity $4 \pi J$. For the Gaunt factor interval this gives time scales from approximately 3.3 to 4.5\,s. The corresponding frequencies are 0.30\,Hz and 0.22\,Hz (in logarithmic values -0.52 and -0.65 respectively). The values are obviously not in agreement with the detected $f_3$ frequency. Therefore, the idea of a fragmented hot accretion flow (conserving the overall initial inhomogeneity at the inner disc) cooled by the free-free emission is not consistent with the detected $f_3$ frequency, unless there is an additional heating source, which suppresses cooling of the boundary-layer material. The most plausible source could be magnetic energy dissipation.

Alternatively, free-free cooling might produce the estimated characteristic frequencies between 0.22 and 0.30\,Hz. In such case, the frequency $f_3$ would be generated by an additional third process, where the fragmentation region could be the disc-corona interaction (see next Section for discussion of the corona). If the disc accretion is turbulent, then the corona accretion can also be highly turbulent. Such unstable accretion flow would be generating unstable X-ray radiation. Therefore, the alternative third scenario could be localised in the corona itself. But any further detailed study is out of scope of this paper.

\subsection{Truncated disc and the inner hot X-ray corona}
\label{corona}


Unfortunately with the present estimate of $\alpha$ and the high uncertainty of measured break frequencies and primary mass, we are not able to determine precisely enough the inner disc radius. We can roughly estimate that the inner disc can reach a maximal radius $0.80 \times 10^9$\,cm. The disc can either be slightly truncated or fully developed down to the stellar surface with an unknown boundary layer radial thickness. For a primary mass of 1.29\,$r_{\rm wd}$, the maximal inner disc radius is about $2.1 \times r_{\rm wd}$. Assuming that the boundary layer radial thickness is correlated with the primary mass (and thus also the primary radius) and the value is $0.4 \times r_{\rm wd}$ (\citealt{narayan1993}), the larger $r_{\rm in}$ values $1.4 - 2.1 \times r_{\rm wd}$ must be generated by a different process. Such disc would be slightly truncated.

The inner disc truncation is generally accepted mainly by theoretical but also by some observational works. It explains the missing boundary layer problem (\citealt{pandel2005}) and the UV delay in dwarf nova outbursts (\citealt{lasota2001}, \citealt{schreiber2003}). Simulations of such outburst activity in SS\,Cyg used an inner disc radius of about $2 \times 10^9$\,cm, which is approximately $5 \times r_{\rm wd}$ (\citealt{schreiber2003}), which is considerably higher than derived in this paper. Optical fast variability known as flickering has been studied and the results also suggest an inner disc truncation in the nova-like star KR\,Aur and the symbiotic system T\,CrB (\citealt{dobrotka2010}, \citealt{dobrotka2012}). The KR\,Aur case gives the inner disc radius in the interval $1.56 - 1.67 \times 10^9$\,cm, which is $1.79 - 1.92 \times r_{\rm wd}$. The white dwarf radii units give similar results to our results for RU\,Peg. The derived T\,CrB inner disc is at radius of about $4 \times 10^9$\,cm which is $20 \times r_{\rm wd}$. The latter value is considerably larger than for RU\,Peg which can be explained by the larger mass of the white dwarf, implying a smaller radius. A truncated disc with $r_{\rm in} = 4 \times r_{\rm wd}$ ($2.5 \times 10^9$\,cm with a 0.91\,M$_{\rm \odot}$ primary from \citealt{catalan1995}) is suggested by eclipse fitting in HT\,Cas during the rise to outburst (\citealt{ioannou1999}). Even though this was not during the quiescence and the inner disc should start to shrink, the values are larger than the estimated RU\,Peg inner disc radius. Furthermore, an important method to study the disc is the eclipse mapping in high-inclination systems. In such study of V2051\,Oph (\citealt{baptista2004}) and V4140\,Sgr (\citealt{borges2005}) the inner disc hole can be estimated from the innermost point in the radial brightness temperature distribution figures. In V2015\,Oph case the disc seems not to be truncated, while in V4140\,Sgr this gives an inner disc radius rough estimate of about $2 \times r_{\rm wd}$ ($1.39 \times 10^9$\,cm) which is consistent with our finding in white dwarf units. However, it is worth to note, that the eclipse mapping resolution is very limited, which can pose problems in identifying whether the disc is truncated or not.

\begin{figure}
\includegraphics[width=85mm,angle=0]{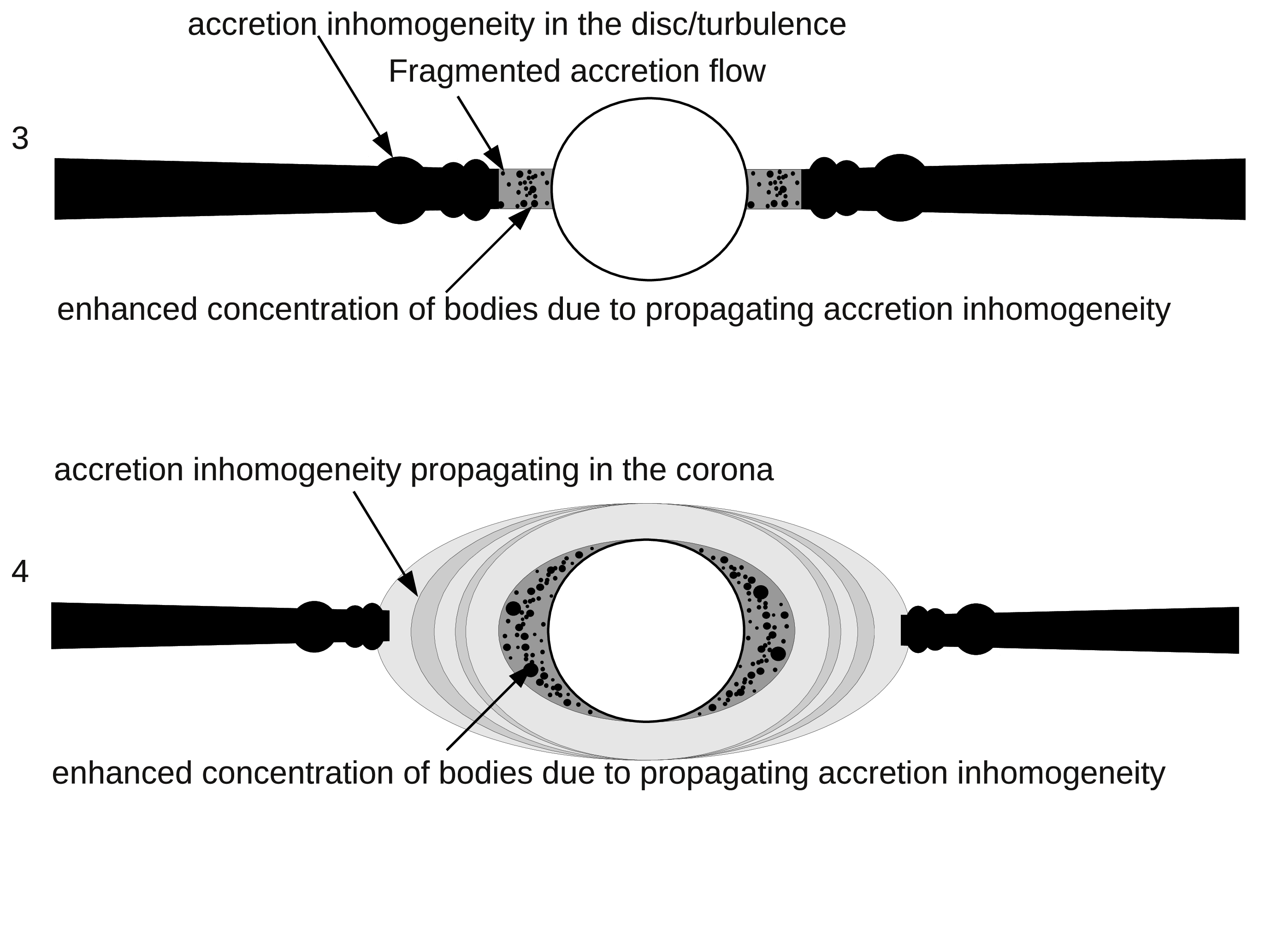}
\caption{Schematic (not to scale) pictures of investigated/discussed models. See text for details.}
\label{ilustration_2}
\end{figure}

An attractive explanation for the inner disc truncation is an evaporated X-ray corona (see e.g. \citealt{meyer1994}). The matter is evaporating from the disc to a hot optically thin corona and every inner disc inhomogeneity is propagating further (situation 2 in Fig~\ref{ilustration_1}). At some point all matter is evaporated and the disc disappears in the optically thick geometrically thin form (inner disc radius). The matter flows through this extended corona toward the central white dwarf, where it interacts with the slowly rotating stellar surface. The tangential velocity starts to decrease near the stellar surface and the kinetic energy is released. This forms the classical boundary layer around the white dwarf with differential rotation fragmenting the matter via Kelvin-Helmholtz instability (situation 4 in Fig.~\ref{ilustration_2}). In order to satisfy the conditions for Kelvin-Helmholtz instability we need two regions with different angular velocity or a fluid with a gradient of this velocity. Following \citet{liu1995}, the matter evaporating from the innermost disc region forms a rotating gas cloud and fills the space between the disc and white dwarf by a geometrically thick tenuous "accretion disc". The evaporated gas has considerable amount of angular momentum and after passing through a turbulent region (due to studied Kelvin-Helmholtz instability) this angular momentum is completely lost, and the gas becomes subsonic. The rotationally supported gas "changes" into pressure supported gas around the white dwarf. Therefore, the idea is the same as the original concept of disc-white dwarf interaction through the boundary layer as a region with strong angular velocity gradient.

Therefore, the possibility that the disc is truncated and that the X-ray corona is fragmenting at the boundary layer outer edge is consistent with recent theories and concepts. More detailed investigation is needed to deduce whether or not the disc is truncated. An independent $\alpha$ parameter determination would help. Finally, the possible maximal inner disc truncation in RU\,Peg is in agreement with few systems only in white dwarf radius units. Otherwise, in cm units the inner disc radius in RU\,Peg is considerably smaller than in other studied systems.

\subsection{Comparison to other observed PDS}
\label{observ_comparison}

\citet{revnivtsev2010} claims that the PDS of optical and X-ray observations of intermediate polars should be similar, because in both cases the flux variability is caused by the same fluctuations of the mass accretion rate. An alternative scenario is reprocessing of X-rays by the disc and re-radiation of the same variability in UV. Furthermore, they suggest that this can be valid also for dwarf novae in the quiescent state. This idea is exactly the same we used in this paper. Following our work, low frequencies satisfy this idea while high frequencies do not.

Imagine a turbulent eddy as a fluctuation of the mass transfer rate at the inner disc with radius of $2.0 \times r_{\rm wd}$ in RU\,Peg system. Then we assume that this blob of matter with a dimension scale of 24\,\% the scale height of the disc will penetrate into the boundary layer with a viscous velocity at the inner radius. Using equations (\ref{Hdisc}), (\ref{cs}) and (\ref{visc_time_scale}) this event will last approximately 3400\,s. This time scale is equivalent to $f_1$ frequency logarithm of -3.53. If we want to get 10 times shorter duration of the event, i.e. frequency logarithm equal to -2.53, we need to account for an accretion eddy with 10 times smaller dimension scale. Approximating the penetrating bodies as spheres, 10 times smaller spheres have 1000 times lower volumes. With the same density, the transferred mass and consequently the released energy is also 1000 times smaller. Such a short event would be negligible compared to the longer ones. This concept is demonstrated in the Fig.~\ref{lc_simulated_visc}. The short events are present following the distribution function (\ref{exp_funkcia_2}), but their amplitudes are so small that the long events dominate.

Therefore, an additional mechanism should produce even fainter fragmentation of the large inhomogeneities. The Kelvin-Helmholtz instability is a promising approach. It can be imagined as spilling a bucket of water out of a window (the accretion fluctuation at the inner disc). The same quantity of water will fall on the floor (large event), but fragmented into small drops (superposed short events). In conclusion, the extrapolation of the intermediate polar picture to dwarf novae presented by \citet{revnivtsev2010} is not certain.

Furthermore, the UV and X-ray PDS coming from different regions should have different characteristics, mainly in the break frequency. We discussed this in Section~\ref{lower_break_frequency}. Similar different UV and X-ray PDS is possible also in the case of SS\,Cyg in quiescence where the PDS in X-rays shows\footnote{depending again on chosen error interval as we did in Section~\ref{pds_uncertainty}} higher break frequency than in UV (\citealt{balman2012}).

Moreover, the dwarf nova VW\,Hyi in quiescence displays at first sight similar PDS in both discussed wavelengths (\citealt{balman2012}). But the X-ray case has an interesting shape. It it very similar to our multicomponent behaviour, i.e. two possible break frequencies, two power laws and two almost constant intervals. The UV break frequency agrees well with the lower break frequency in the X-ray PDS. This suggests the same behaviour as discussed in the previous paragraphs, where the inner disc generates larger fluctuations (displayed as break frequency in UV PDS and lower break frequency in X-ray PDS) which are subsequently fragmented into smaller bodies and observed as fast X-ray variability (displayed as higher break frequency in X-ray PDS). But the observed X-ray PDS of VW\,Hyi in \citet{balman2012} has large errors in high frequencies, hence our interpretation is questionable.

Our concept of low and high break frequencies arising in the accretion disc and boundary layers, respectively is not unique to RU\,Peg alone as multicomponent power spectra were also observed in X-ray binaries. \citet{sunyaev2000} summarised PDS characteristics of 9 black hole and 9 neutron star X-ray binaries in quiescence. Some power spectra from neutron star systems show similar characteristics as we found for RU Peg, i.e. a multicomponent shape with four different components and two characteristic break frequencies (see upper panel of Fig.~\ref{pds_observed}). The most significant similarity can be seen for the systems 4U1608-522, GX354-0 and KS 1731-260. 4U1705-44 has also the same shape, but in this case the low frequency part is not a straight line. The PDS of 4U0614+091 is also similar but with an additional broken power law (break frequency). Following the authors, the lower break frequency can be generated in the inner disc edge caused by the viscous process (see also \citealt{done2007}), and the high frequency part is generated in the boundary layer near/at the surface of the neutron star. \citet{sunyaev2000} propose different physical scenarios to explain the high frequency variability. They also discuss Kelvin-Helmholtz instabilities, differently than we do, but their explanation applies better to periodic or quasiperiodic events. Our model is an alternative that can potentially aid future detailed studies of neutron star systems. Finally, owing to the absence of a solid surface in black hole systems, none of the PDS show a multicomponent shape.

\section{Summary}
\label{summary}

In this paper we study fast X-ray variability of the dwarf nova RU\,Peg in quiescence. We present a new X-ray analysis of data and elucidate the statistical "toy model" for light curve simulation based on unstable mass accretion onto the white dwarf. By analysing the simulated light curves we searched for observed PDS parameters. Our findings are the following:

(i) The PDS analysis of X-ray data exhibits more complicated structure than a simple broken power law. It is composed of two broken power laws.

(ii) It is known that the X-rays in cataclysmic binaries are produced by the interaction of the accreted matter with the central white dwarf in the so-called boundary layer or in the X-ray corona. The non steady radiation is produced by the non steady accretion flow. Unstable mass supply from the turbulent inner disc explains the low frequency part of the observed PDS. Additional processes between the inner disc and the white dwarf must generate the observed fast variability patterns.

(iii) If the disc is truncated, the truncation radius is no larger then $0.80 \times 10^9$\,cm. This value is smaller compared to the other studied cataclysmic variables in quiescence. A fully developed disc with an unknown boundary layer radial thickness can not entirely be excluded, owing to the large uncertainties in the measured break frequencies, primary mass and the large range of studied values of the disc parameter alpha.

(iv) Any fluctuation in mass accretion at the inner accretion disc which is supplying the corona or boundary layer is manifesting in UV and X-rays in the inner disc and corona/boundary layer, respectively. This explains the low frequency break of the PDS to be identical in UV and X-rays. Further fragmentation of the flow produces variability on shorter time scales, visible only in X-rays and characterised by a second higher frequency break in the X-ray PDS.

(v) The multicomponent shape of the PDS where the lower break frequency is generated in the accretion disc while the high break frequency by the boundary layer is not unique for RU\,Peg dwarf nova system. Observations of similar PDS in some neutron star X-ray binaries yield similar interpretation. The black hole binaries do not show this PDS shape probably due to the absence of solid surface.

While our model approach follows principles of plausibility, we have no ultimate proof that it reflects reality. Any future superior models may modify the conclusions about inner disc radius and the disc truncation, as these are particular dependent on the underlying model assumptions.

The fast X-ray variability remains unresolved. We can just conclude that it is probably generated in the corona or the boundary layer. Further X-ray observation with higher time resolution is required to test whether the PDS has more components than observed. The free-free cooling of the boundary layer plasma suggests a possible break frequency of about -0.52 to -0.65 in logarithmic values. If the free-free cooling is affected by an additional heating process, this can move the estimated break frequencies to the observed fast variability values. Furthermore, unstable turbulent accretion flow in the corona can also produce unstable X-ray radiation. The disc-corona interaction can be also highly turbulent. Therefore, further modelling of the coronal flow is required to understand the fast X-ray variability in this and similar objects exhibiting multicomponent PDSs.

\section*{Acknowledgements}

We thank the anonymous referee for very constructive report which yields considerable improvement of this paper. AD was supported by the Slovak Academy of Sciences Grant No. 1/0511/13, by the ESA international fellowship and by the Japanese Society for Promotion of Science (JSPS). SM was supported in part by the Grant-in-Aid of Ministry of Education, Culture, Sports, Science, and Technology (MEXT) (22340045, SM) and by the Grant-in-Aid for the global COE programs on The Next Generation of Physics, Spun from Diversity and Emergence from MEXT. We are grateful to the HPC centre at the Slovak University of Technology in Bratislava, which is a part of the Slovak Infrastructure of High Performance Computing (SIVVP project, ITMS code 26230120002, funded by the European region development funds, ERDF), for the computational time and resources made available.

\bibliographystyle{mn2e}
\bibliography{mybib}

\label{lastpage}

\end{document}